\newif\ifisanonymous

\isanonymousfalse

\newcommand{\revivd}{\ifisanonymous VRT\else ReViVD\fi}

\newcommand{\city}{\ifisanonymous a European city\else Lyon, FR\fi}

\newcommand{\gps}{\textsc{Rugby}}
\newcommand{\propeller}{\textsc{Propeller}}
\newcommand{\particles}{\textsc{Particles}}
\newcommand{\traffic}{\textsc{Traffic}}

\newcommand{\nbExperts}{6~}

\documentclass{vgtc}                          

\ifpdf
  \pdfoutput=1\relax                   
  \usepackage{graphicx}                
  \DeclareGraphicsExtensions{.pdf,.png,.jpg,.jpeg} 
\else
  \usepackage{graphicx}                
  \DeclareGraphicsExtensions{.eps}     
\fi%

\graphicspath{{figures/}{pictures/}{images/}{./}} 

\usepackage{microtype}                 
\PassOptionsToPackage{warn}{textcomp}  
\usepackage{textcomp}                  
\usepackage{mathptmx}                  
\usepackage{times}                     
\usepackage{cite}                      
\usepackage{tabu}                      
\usepackage{booktabs}                  

\usepackage{xspace}
\newcommand{\ie}{i.\,e.\xspace}

\newcommand{\eg}{e.\,g.,\xspace}

\onlineid{1681}

\vgtccategory{Research}

\vgtcinsertpkg

\title{\revivd: Exploration and Filtering of Trajectories in an Immersive Environment using 3D Shapes}

\author{François Homps\thanks{e-mail: francois.homps@ecl17.ec-lyon.fr} %
\and Yohan Beugin\thanks{e-mail: yohan.beugin@ecl17.ec-lyon.fr} %
\and Romain Vuillemot\thanks{e-mail: romain.vuillemot@ec-lyon.fr}}
\affiliation{\scriptsize LIRIS, \'Ecole Centrale de Lyon, France}

\teaser{
\centering
\includegraphics[width=.24\textwidth]{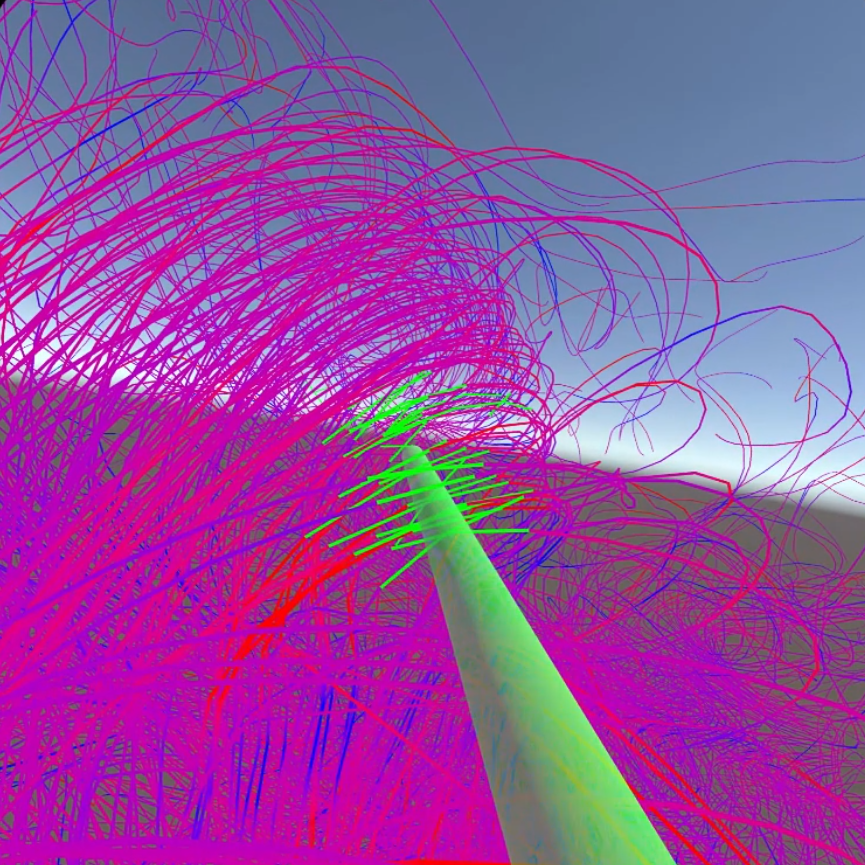}
\includegraphics[width=.24\textwidth]{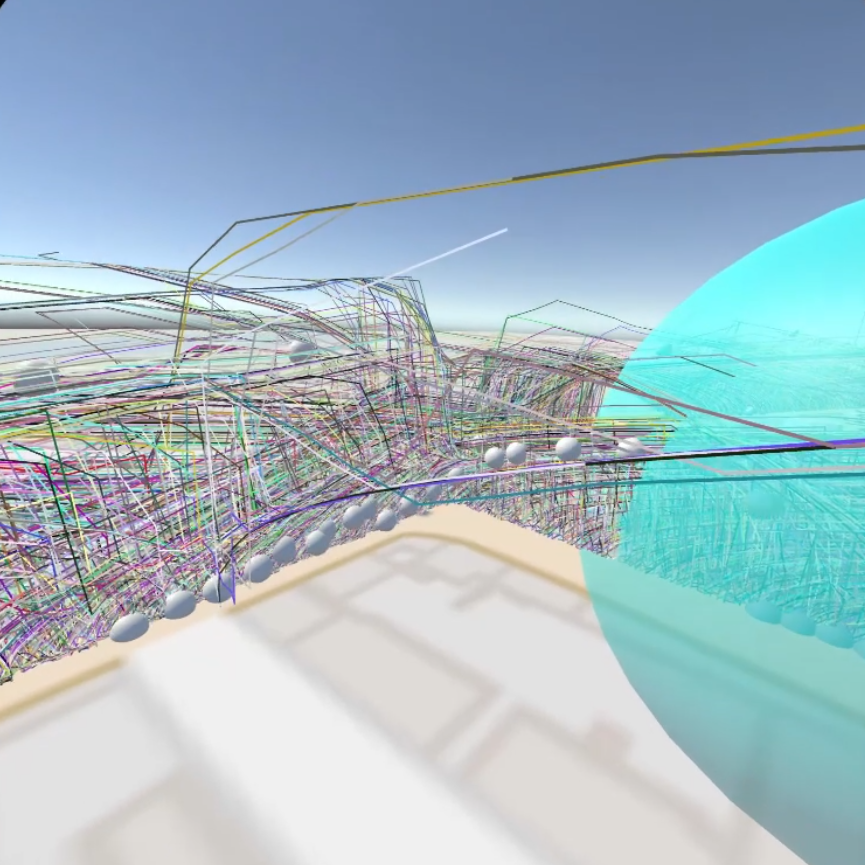}
\includegraphics[width=.24\textwidth]{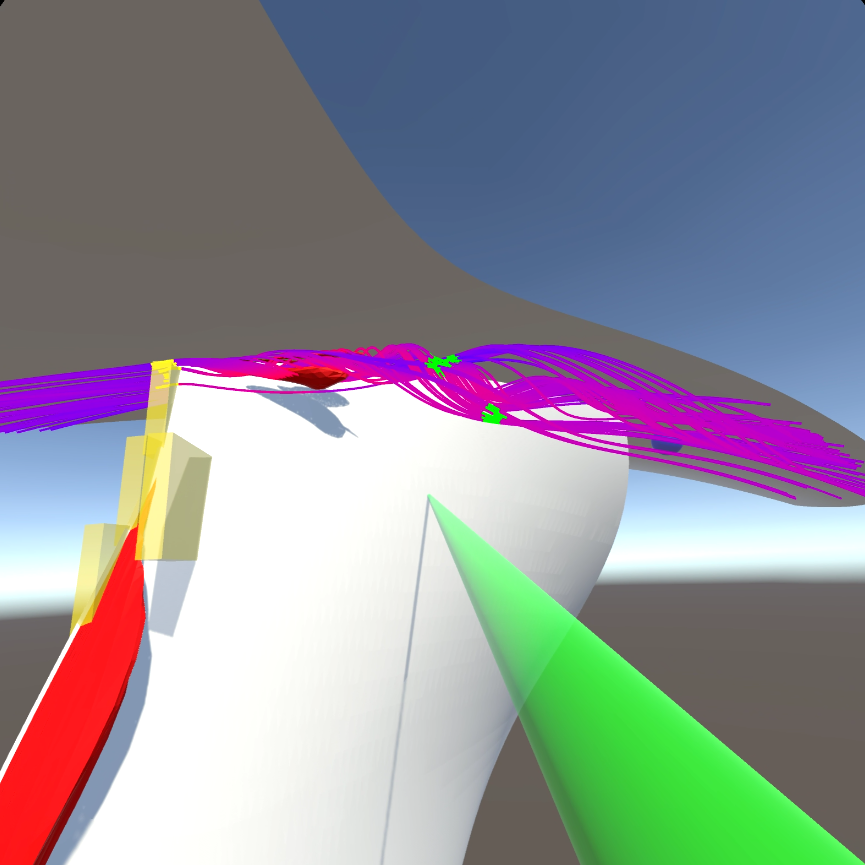}
\includegraphics[width=.24\textwidth]{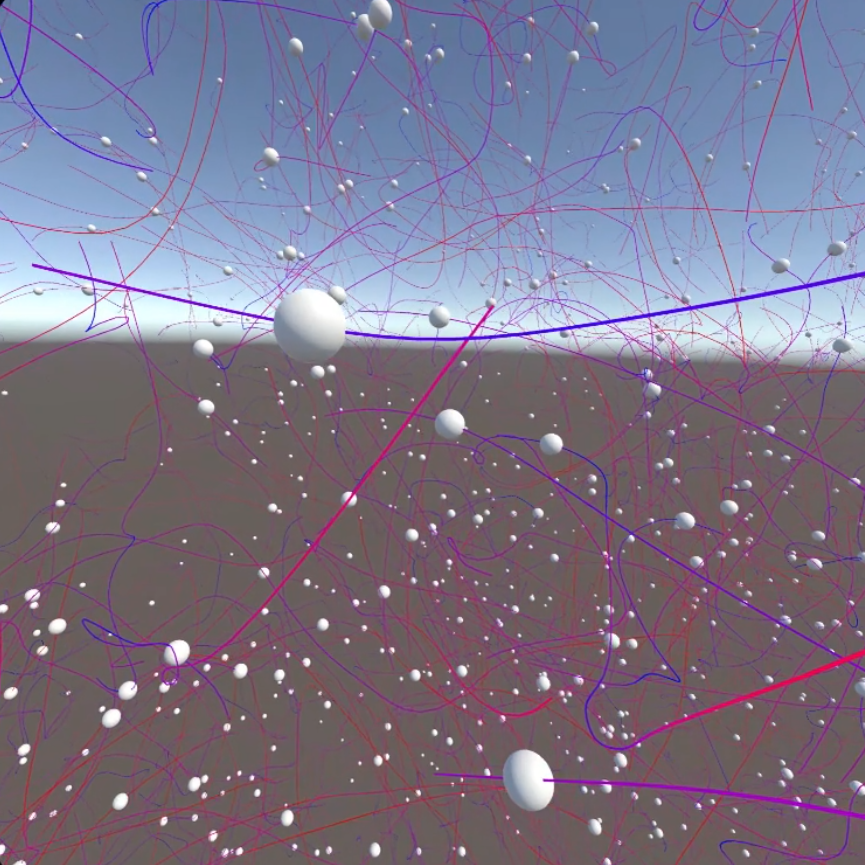}
\caption{\revivd~enables dynamic filtering of large trajectory datasets using 3D shapes. From left to right: (a) A \emph{cylinder} used as a saber to select and filter particles from a numerical simulation dataset, (b) a \emph{sphere} used to select and animate vehicles at a road junction in a car traffic dataset. Such shapes can become (c) \emph{persistent} to refine queries and extract complex vortex, and trajectories can be (d) \emph{animated} to re-create turbulent flows of particles.}
\label{fig:teaser}
}

\abstract{
We present \revivd , a tool for exploring and filtering large trajectory-based datasets using virtual reality. \revivd 's novelty lies in using simple 3D shapes\textemdash such as cuboids, spheres and cylinders\textemdash as queries for users to select and filter groups of trajectories. Building on this simple paradigm, more complex queries can be created by combining previously made selection groups through a system of user-created Boolean operations. We demonstrate the use of \revivd\ in different application domains, from GPS position tracking to simulated data (\eg turbulent particle flows and traffic simulation). Our results show the ease of use and expressiveness of the 3D geometric shapes in a broad range of exploratory tasks. \revivd\ was found to be particularly useful for progressively refining selections to isolate outlying behaviors. It also acts as a powerful communication tool for conveying the structure of normally abstract datasets to an audience.
} 

\CCScatlist{
  \CCScatTwelve{Human-centered computing}{Human computer interaction (HCI)}{Interaction paradigms}{Virtual Reality};
  \CCScatTwelve{Human-centered computing}{Visu\-al\-iza\-tion}{Visualization systems and tools}{Visualization toolkits}
}

\begin{document}

\firstsection{Introduction}

\maketitle

Many fields generate large trajectory datasets: GPS trackers on aircraft are used for real-time monitoring of thousands of moving objects; road traffic simulations give insights on how to prevent traffic jams in a city; particle simulations of turbulent flows make it possible to identify phenomenons normally invisible due to their small scale in space and time. Those domains usually involve domain-expert users who need to explore their dataset to assess the structure and quality of collected or generated data~\cite{tukey_exploratory_1977}. 

Interaction is key during the exploration process, in particular to isolate elements of interest. For instance, when interacting with aircraft datasets, a frequent exploratory task is to filter by take offs and landings, for a given specific airport area. Assuming the planes' positions are displayed in 3D with coordinates (x, y, z), and their trajectories result from their tracking over time (t), selecting planes is achieved by having constraints on both 1) the 2D position on the ground (a circle centered on the airport), and 2) maximal altitude. Performing such a selection is difficult with standard 2D dynamic queries~\cite{shneiderman_dynamic_1994} as they only operate on a plane. 

Immersive environments demonstrate great promise to explore and interact with 3D datasets~\cite{chandler_immersive_2015}. Recent work includes Fiberclay~\cite{hurter_fiberclay:_2019}, which provides a tilted 3D view, direct brushing and selection interactions. However, those selections are not localized and also select fly-overs; the process requires deselecting trajectories from another point of view.

This paper seeks to introduce a simple yet general-purpose interaction technique that operates simultaneously on the three spatial dimensions. The technique uses 3D geometric shapes (spheres, cuboids and cylinders) that encode a 3D query which the user manipulates to probe a highly cluttered environment~\cite{elmqvist_taxonomy_2008}. Our research hypothesis is that such 3D geometric queries are not only expressive enough for open-ended exploration tasks, but also to single out specific elements. To validate this hypothesis, we present \revivd \ifisanonymous ~[name changed for peer review]\fi, a VR trajectory data visualizer in which we implemented the technique and tested it against different application domains and datasets. Our paper has three main contributions:

\begin{itemize}
    \item The first contribution is to present a visualization toolkit that allows non-experts in Virtual Reality  to intuitively create and use simple 3D geometric shapes as selection tools, and to filter the dataset through the use of custom Boolean operations.
    \item The second contribution of this paper is the implementation of persistent selectors that allow the reuse of specific queries.
    \item The third contribution is the mapping choice of all the interactions possible in \revivd~on the limited inputs available with the VR controllers. 
\end{itemize}

\section{Related work}

\subsection{Visualization in Immersive Environments}

Immersive environments (\eg VR, augmented reality, interactive walls) have demonstrated great promise for the exploration of abstract datasets~\cite{chandler_immersive_2015}. In particular, recent years have seen a fast democratization of Virtual Reality (VR) technology: newer headsets with much more resolution and precise user's movements, VR support in consumer-grade graphics card and the emergence of high-level development software such as Unity\footnote{\url{https://unity.com/}}, along with standardized libraries like Valve's OpenVR\footnote{\url{https://github.com/ValveSoftware/openvr}}, have made standardized VR technology accessible. 

In the meantime, the task of exploring large datasets has become pervasive in all research and engineering domains. We found that while extremely powerful visualization tools, such as Kitware's ParaView\footnote{\url{https://www.paraview.org/}} and its plugins like the Topology Toolkit~\cite{tierny_topology_2018} exist, they lack the ease of use and intuitiveness that VR shines in.

\ifisanonymous
This is why we started work on \revivd, a lightweight open-source tool aimed at science and engineering domain-experts specialized in quickly visualizing and filtering trajectory datasets in VR.
\else
This is why we started work on \revivd\ (from \textit{Réalité Virtuelle~: Visualisation de Données}, French for data visualization in virtual reality), a lightweight open-source tool aimed at science and engineering domain-experts specialized in quickly visualizing and filtering trajectory datasets in VR.
\fi


Indeed, when compared to traditional 3D visualizers, VR shows high potential. Users can immerse themselves in the dataset with many more degrees of freedom than through a traditional mouse and keyboard interface, both for the control of their point of view and of their interactions with the data. However, these benefits come at a cost; depth perception and  comfortable viewing are only possible on dual high resolution screens, which greatly impact the rendering performance. Meanwhile, a high frame rate (traditionally, a minimum of 30 fps) is essential in VR to reduce risks of nausea. 

\subsection{Exploration in Immersive Environments}

Exploration techniques in those settings often rely on displaying large amounts of data for the user to navigate within, with no previous assumptions (\eg filtering, aggregation) made. Fiberclay~\cite{hurter_fiberclay:_2019} is a flagship example that displays thousands of trajectories and lets the user build queries using beams. ImAxes~\cite{cordeil_imaxes:_2017} and IATK \cite{cordeil_iatk:_2019} permit the creation of axes to map data attributes to spatial locations and create standard charts (\eg parallel coordinates, scatterplots) based on logical rules. Those techniques have in common that they require no user interface to switch between interaction modes: only the users' VR controllers are visible in the virtual environment. \revivd\ is heavily inspired by this approach and augments the controllers to display current possible queries as 3D shapes called selectors, as well as operation menus used to filter the dataset by combining multiple selections.

\subsection{Dynamic Queries and Brushing}
Dynamic queries~\cite{shneiderman_dynamic_1994} set the foundation for interactive exploration of large datasets. They enable the selection of value ranges in an iterative and safe way and are informative as they are coordinated to visual feedback. Such queries enable users to grasp the structure and distribution of the datasets through transient selections, 
 which are key for exploratory data analysis~\cite{tukey_exploratory_1977}. Our work is built on this paradigm, and in particular \emph{brushing}, which complies with the direct manipulation paradigm to let users manipulate regions of interest on the screen  in a natural way. Brushing has been extended beyond 2D environments such as in 3D~\cite{sherbondy_exploring_2005} and  immersive environments, \eg for Cave interactions~\cite{symanzik_dynamic_1996}, in Virtual  Reality~\cite{hurter_fiberclay:_2019, prouzeau_scaptics_2019}, or with tangible  interactions~\cite{besancon_hybrid_2019, jackson_lightweight_2013}. However, brushing specific areas or volumes remains  challenging especially with large datasets due to occlusions.

\subsection{3D Occlusions and Selection Tools}

3D environments require occlusion management~\cite{elmqvist_taxonomy_2008}, as objects in a scene can mask each other, especially when their density is important. To facilitate large selections, different techniques have been used in VR such as hand avatars, bimanual selection, 3D cursors and raycasting \cite{argelaguet_survey_2013}. While ray casting allows ease of interaction and selection in Virtual Reality and seems to have become a default technique \cite{argelaguet_survey_2013}, it does not allow for distinguishing between objects that are both on the infinite ray, except for adding a cursor along the ray to select the closest object \cite{baloup_raycursor:_2019}. Other techniques such as shapes (\eg squares, circles) can be used to select all items \emph{within} them. In a 2D space, rectangle shapes are standard as they can easily be built using mouse selection\cite{shneiderman_dynamic_1994,van_den_elzen_multivariate_2014}, but if more attributes are being used, complex shapes can emerge~\cite{heer_generalized_2008}, \eg by using a non-visual attribute. In 3D worlds, 3D shapes (\eg cuboid, sphere) enable the manipulation of spatial regions in an efficient way to select groups of objects and so progressively refine the selection \cite{kopper_rapid_2011, jackson_force_2012}. In 3D point clouds, context-aware selections based on heuristics and machine learning intention filtering have been proven useful to perform better selections \cite{chen_lassonet:_2019, yu_cast:_2016}. BalloonProbe~\cite{elmqvist_balloonprobe:_2005} uses spherical shapes to let users separate objects based on their attributes in a 3D scene. Bounding boxes of groups of points in a 3D space~\cite{hentschel_virtual_2009} can become interactive to select items. Balloon Selection~\cite{benko_balloon_2007} also lets users create and translate spheres in a 3D space for target selections, with a combination of 2 DOF interactions. 
Flat 2D planes can be used for spatial structure projection, \eg in brain imagery~\cite{chen_human_2011, akers_cinch:_2006, coffey_interactive_2012}. 

To alleviate the problem of occlusion, already lessened by the small footprint of the trajectories, our work capitalizes on using a variety of probes both in shapes and sizes, as well as by providing an ease of navigation to make better points of view quickly available to the user.


\begin{figure*}[t!]
    \centering
    \includegraphics[width=.50\columnwidth]{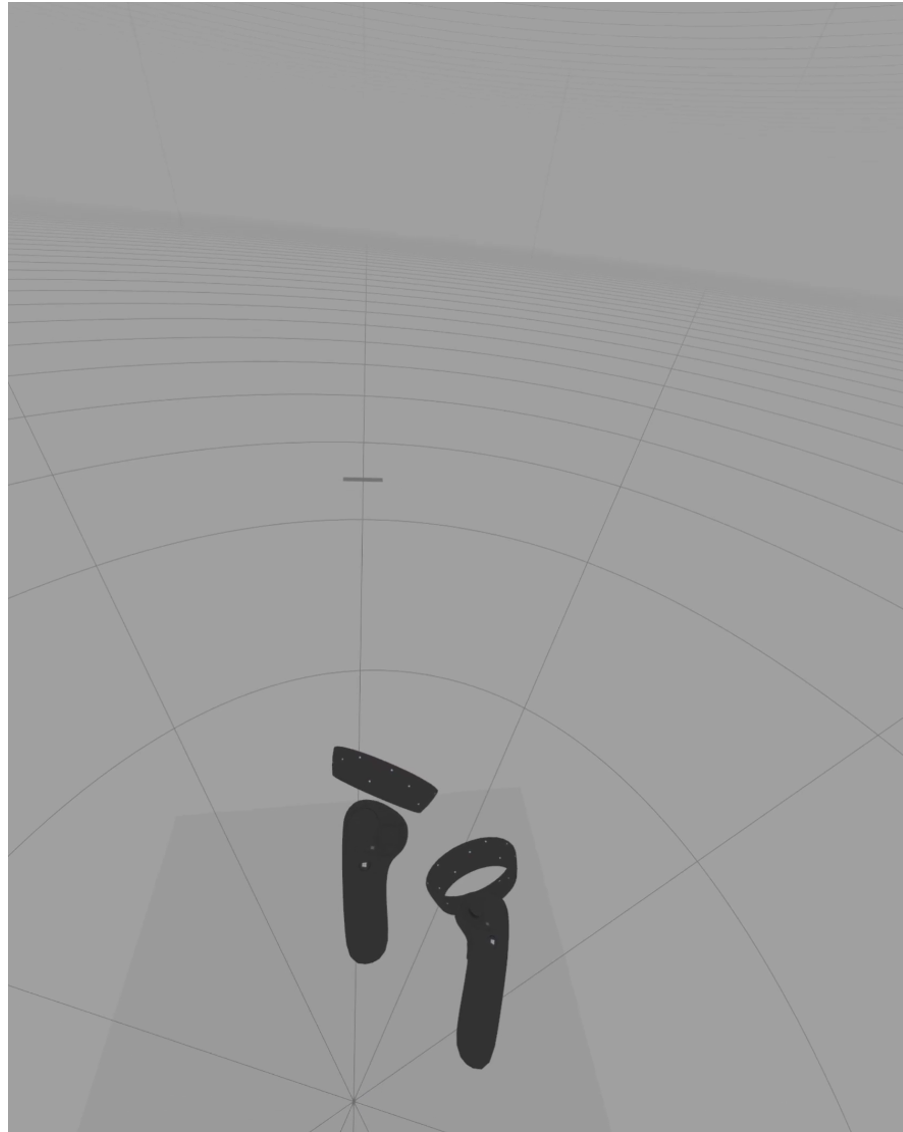}
    \includegraphics[width=.50\columnwidth]{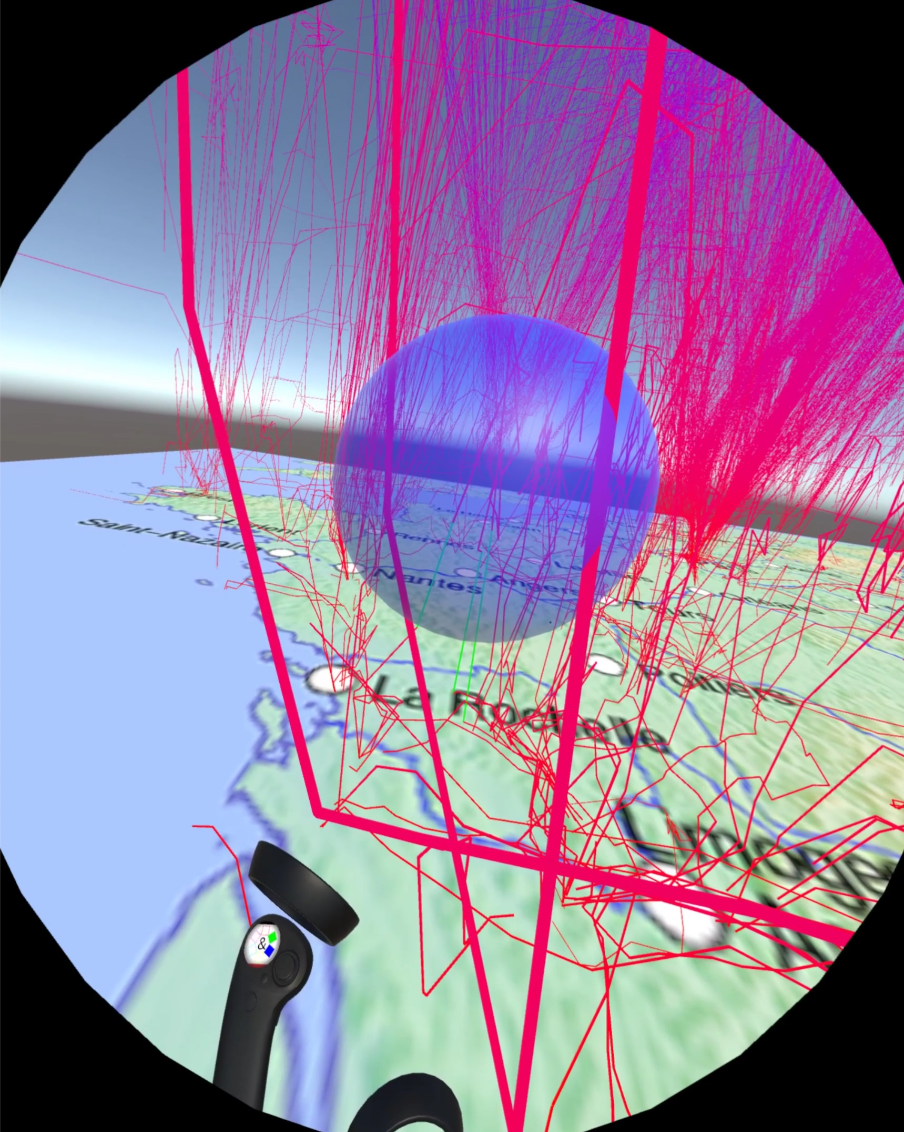}
    \includegraphics[width=.50\columnwidth]{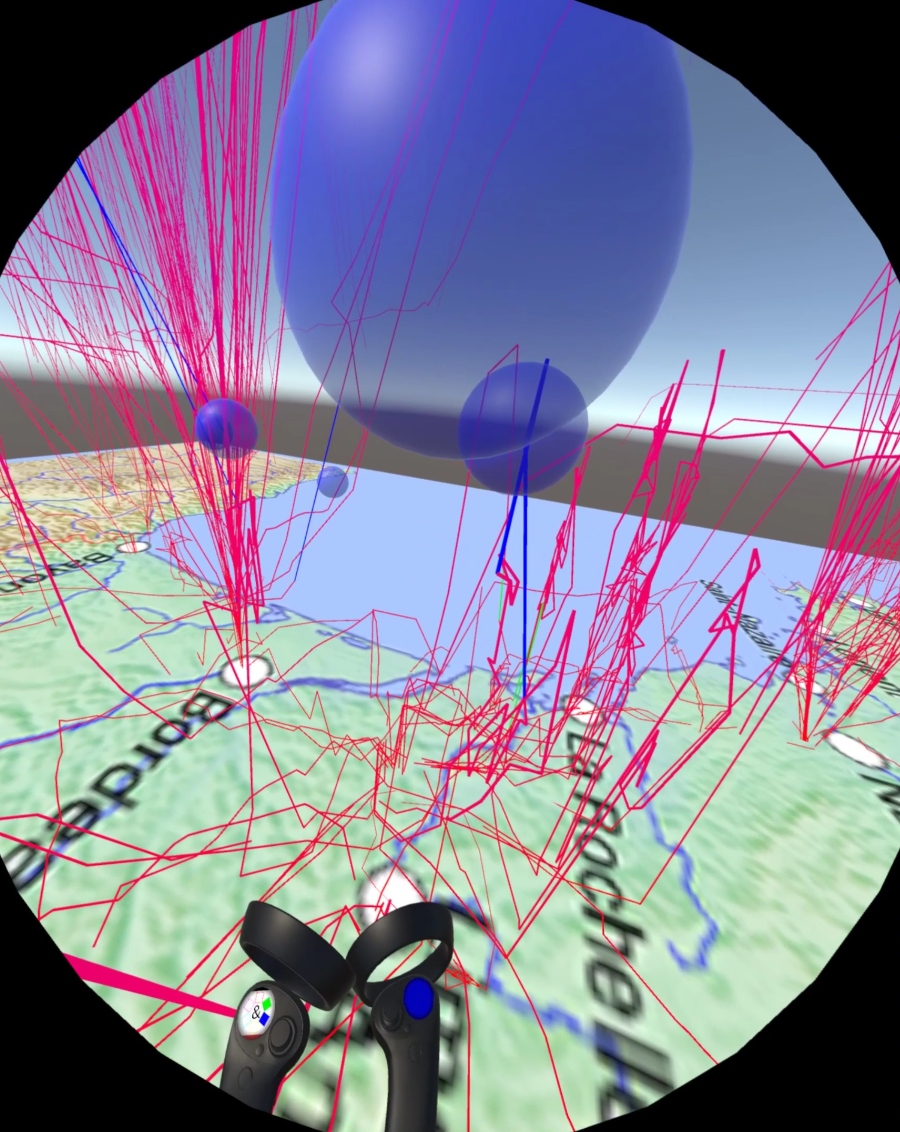}
    \includegraphics[width=.50\columnwidth]{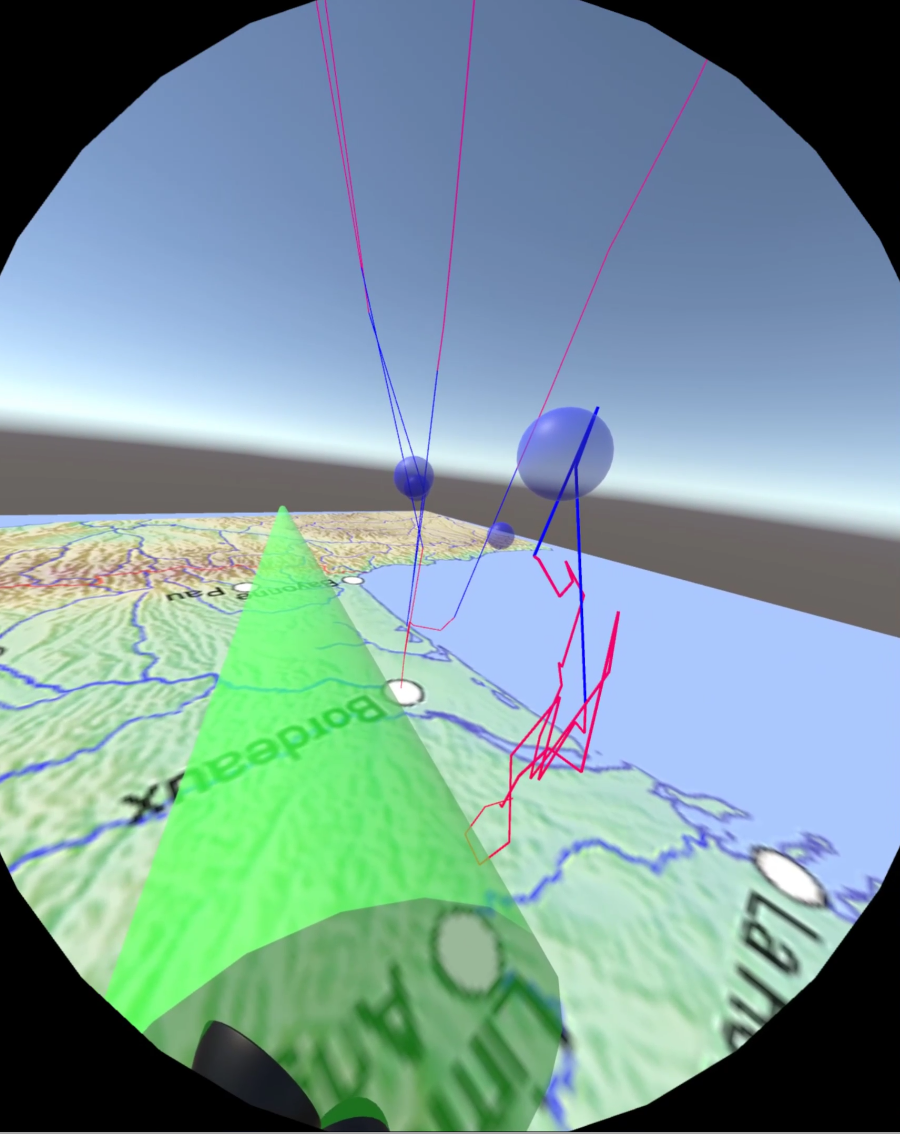}
    \caption{Overview of a standard session of exploration using \revivd. From left to right: (a) SteamVR environment seen while configuring and loading the session; (b) Hand-held selection using a sphere; (c) Persistent selectors used for repeated selections; (d) Filtering trajectories based on previous selections.}
    \label{fig:revivd_intro}
    \vspace{-.3cm}
\end{figure*}

\section{Features Design}
The design approach of \revivd\ is based on three core features illustrated \autoref{fig:revivd_intro}:
\begin{enumerate}
    \item Visualization: \revivd\ allows the user to see a 3D rendering of the raw dataset and easily change their point of view through intuitive game-like movement controls.
    \item Selection: intuitive arm motions allow the user to select the parts of the dataset they want to filter by touching them with configurable 3D shapes called selectors.
    \item Filtering: the user can create custom Boolean operations to control the visibility of the trajectories they selected.
\end{enumerate}

This section details the design of those three features.

\subsection{Immersive Visualization of Trajectory Data}

At the core of \revivd\ is its ability to display trajectory data in VR. The \revivd\ engine renders a visualization as an ensemble of independent \emph{paths}, each made of many small 2D segments called \emph{ribbons}. Each ribbon serves to link two adjacent points of the path. We name each of these points \emph{atoms}, as they can hold more attributes than spatial coordinates (\eg a time value, the current kinetic energy of a particle, etc.). Examples of visualizations can be seen \autoref{fig:teaser}.

Additionally, \revivd\ enables time-based animation of trajectories using small spheres that follow each path from its start to its end (optionally using its atoms' time values for reference if applicable). The position, radius, and animation speed of these spheres can be configured in real time, making them a good tool to expose invisible implicit attributes such as speed. This technique is similar to one used in flow visualizations to visualize particles \cite{engelke_autonomous_2019}.

Trajectories can also be animated from a shared starting position, \ie a given spatial selection from which a playback starts, ignoring the context of the actual starting time of each path. We call this technique "dropping" spheres on a selection.

\subsection{Selection using 3D Geometric Queries: Selectors}

A selector is a tool for selecting trajectories; it takes the form of a colored 3D shape attached to the hand of the user. It functions as a 3D brushing tool, allowing the user to make 3D geometric queries on the dataset. There are six colors available for selectors (red, green, blue, yellow, cyan and magenta). Thus, the user has six stored selectors that they can use by switching from one color to the other.

When the user presses a physical selection trigger (\autoref{fig:input_mapping}), the selector becomes active, and all visible ribbons touching the selector take the color of the selector. This notifies the user that the paths those ribbons belong to have been selected and put in that selector's color group (used afterwards through operations). The selector then stays active until the user lets the trigger go, allowing for selection through painting-like strokes. It is important to note that while only a part of the path is colored, the entire path is considered selected in that color. This partial coloration does not hide too much of the original color of the path, which can be configured to relay information. 

We include three selector shapes in \revivd: \emph{cuboids}, \emph{spheres} and \emph{cylinders} (see \autoref{tab:prim_params}). These shapes can be manipulated through \emph{parameters} which define their size as well as their position and rotation relative to the hand of the user. Controllable parameters, listed \autoref{tab:prim_params}, are intentionally restrictive. This is necessary in order to not confuse the user with too many possible operations on the restricted number of buttons available on the VR controllers. Cuboids have the additional limitation of being rotation-locked on both horizontal axes; this way, their bottom side is always parallel to the ground of the visualization, making axis-aligned queries easier to perform.

There is some redundancy in functionality in this choice of three shapes; notably, a default cube can be used as a long-distance brush in a similar fashion to the sphere. This was a conscious implementation decision, as each of the three shapes are still specialized for particular actions for which they are more intuitive to use.

Spheres are mostly useful for selecting all trajectories passing through a specific point of interest, or used as a detached pointer to select specific trajectories at a medium distance from the user. Cylinders are often used as a kind of sword, slicing through a bundle of similarly-oriented trajectories to select them all. Cuboids, thanks to their rotation lock, are more useful as walls, ceilings or floors delimiting certain regions of space, \eg for selecting all trajectories going above a certain height.

\begin{table}[t]
\centering
\begin{tabular}{c|c|c|c|}
\cline{2-4} & \multicolumn{3}{c|}{\textbf{Parameters relative to the hand of the user}}            \\ \hline
\multicolumn{1}{|c|}{\textbf{Shape}} & \begin{tabular}[c]{@{}c@{}}\textbf{Offset} \end{tabular} & \textbf{Size} & \textbf{Rotation} \\ \hline

\multicolumn{1}{|c|}{ \begin{tabular}[c]{@{}c@{}} \includegraphics[width=1.6cm]{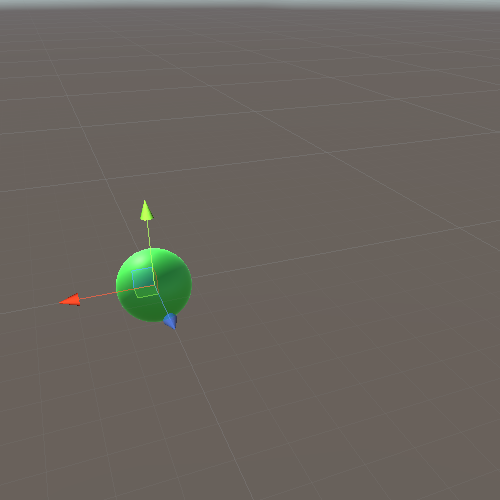}\\ Sphere\\~\end{tabular}}&\begin{tabular}[c]{@{}c@{}} \includegraphics[width=1.6cm]{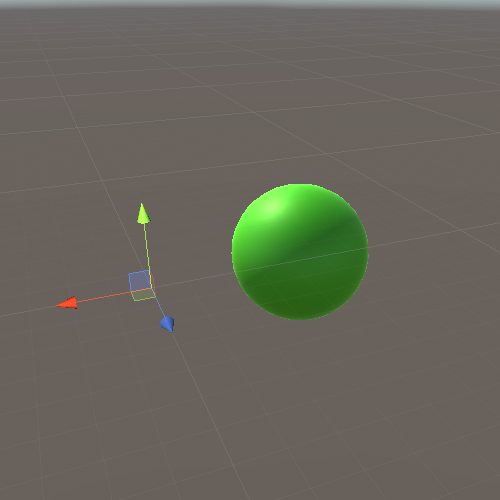}\\ Offset\\ (Vector3)\end{tabular}&\begin{tabular}[c]{@{}c@{}} \includegraphics[width=1.6cm]{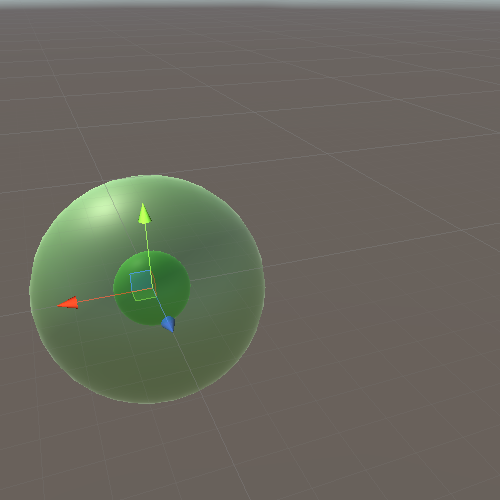}\\ Radius\\ (float)\end{tabular}&    \begin{tabular}[c]{@{}c@{}} \includegraphics[width=1.6cm]{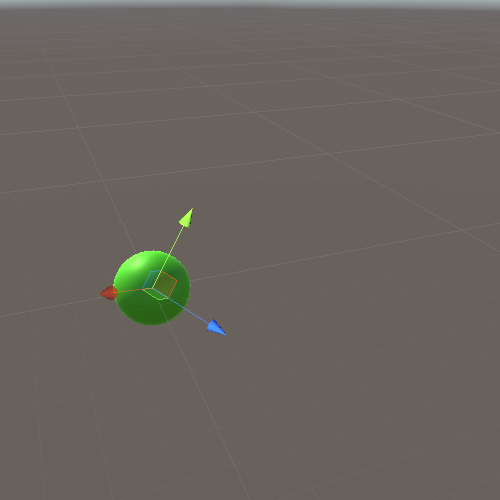}\\ RotXYZ\\ (Vector3)\end{tabular}\\ \hline

\multicolumn{1}{|c|}{ \begin{tabular}[c]{@{}c@{}} \includegraphics[width=1.6cm]{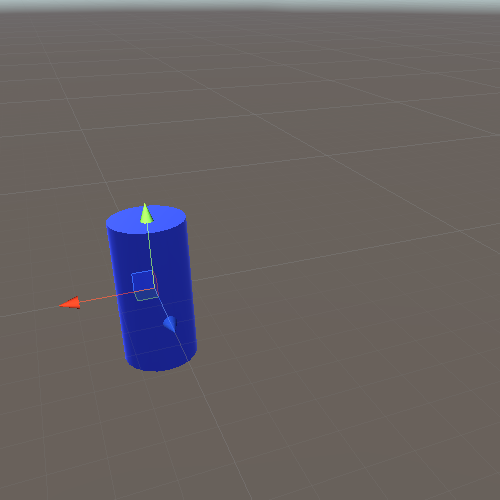}\\ Cylinder\\ ~\end{tabular}}&\begin{tabular}[c]{@{}c@{}} \includegraphics[width=1.6cm]{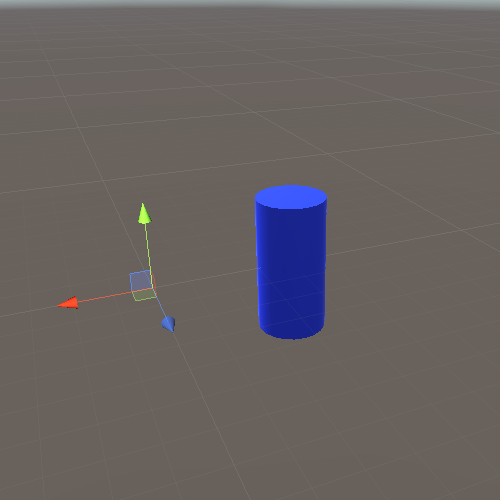}\\ Offset\\ (Vector3)\end{tabular}&\begin{tabular}[c]{@{}c@{}} \includegraphics[width=1.6cm]{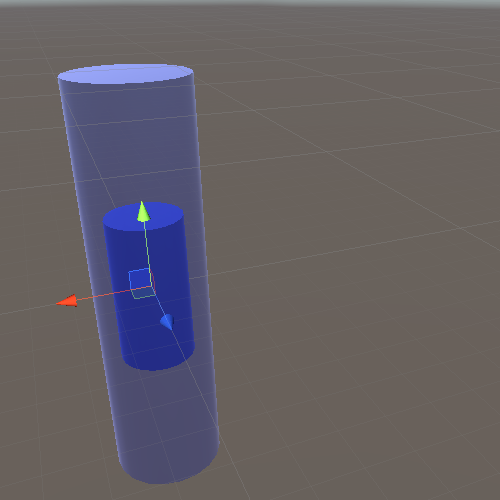}\\ Radius(float) \\ Width (float)\end{tabular}&    \begin{tabular}[c]{@{}c@{}} \includegraphics[width=1.6cm]{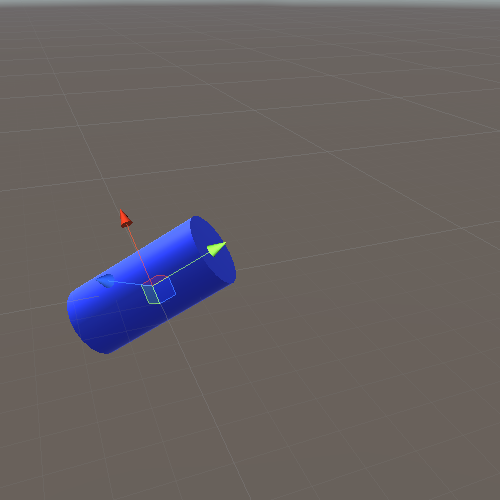}\\ RotXYZ \\ (Vector3)\end{tabular}\\ \hline

\multicolumn{1}{|c|}{ \begin{tabular}[c]{@{}c@{}} \includegraphics[width=1.6cm]{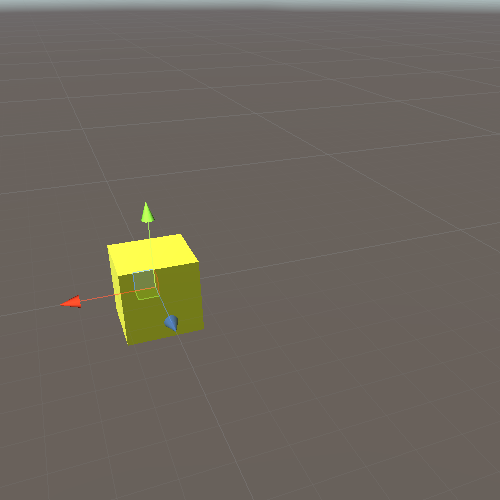}\\ Cuboid\\~\end{tabular}}&\begin{tabular}[c]{@{}c@{}} \includegraphics[width=1.6cm]{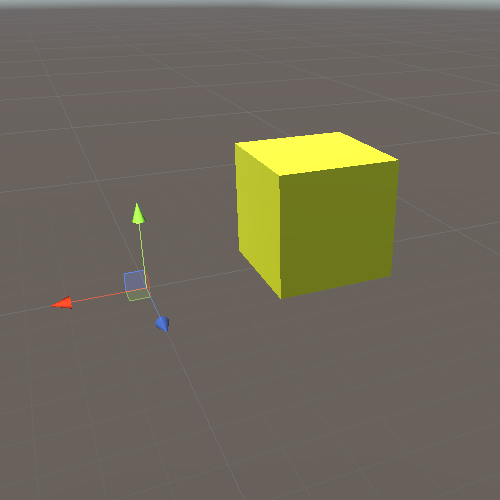}\\ Offset\\ (Vector3)\end{tabular}&\begin{tabular}[c]{@{}c@{}} \includegraphics[width=1.6cm]{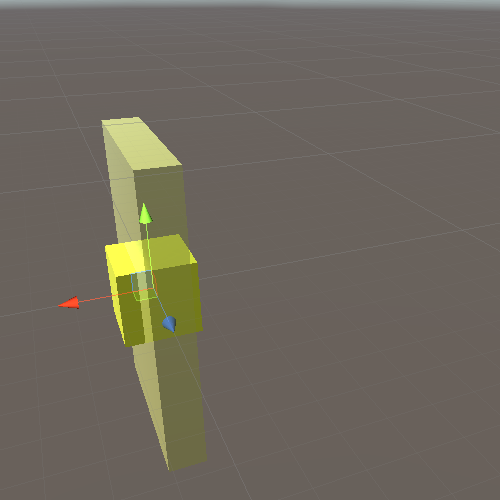}\\ SizeXYZ \\ (Vector3)\end{tabular}&    \begin{tabular}[c]{@{}c@{}} \includegraphics[width=1.6cm]{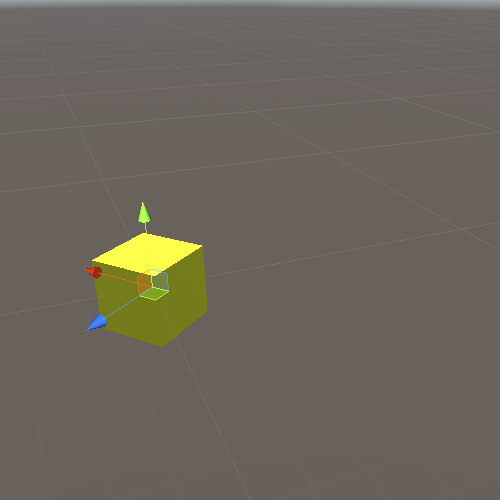}\\ RotZ\\ (float)\end{tabular}\\ \hline

\end{tabular}
\vskip .5cm 
\caption{Selector shapes and parameters implemented in \revivd.}
\vspace{-.5cm}
\label{tab:prim_params}
\end{table}

\subsubsection{Persistent selectors}

It can often be interesting to perform the same query multiple times - for example, when redoing a selection after having hidden a subset of ribbons. In this case, using hand-held shapes to try to make the exact same query is cumbersome. To solve this problem, we introduce \emph{persistent} selectors: by pressing a button, the user can clone the currently hand-held selector in the 3D space as a persistent selector, which won't move with the user and can be activated at a distance when necessary.

Persistent selectors also give the user the opportunity to seed multiple points of interest ahead of time, to use later in combined queries - this is especially interesting in fields focused on such points of interest, \eg with traffic data.

While all three selector shapes can be placed as persistent selectors, they are mostly intended for use with spheres and cuboids, as cylinder-type selectors are often more useful for hand selection.

\subsection{Using Selections to Filter Data: Operations}

Once selections have been done, the user can utilize the color groups they populated to filter the currently displayed trajectories. This process of filtering is done through the use of operations that the user can create on the fly. Operations are characterized by their \emph{type} (OR or AND) and the color groups they hold (which can be any combination of the six selector colors).

An OR-type operation will, out of all currently displayed paths, hide all but those that are selected in at least one of the operation's color groups; as an example, the "Blue, Green, OR" operation will only keep displayed the trajectories that are either selected in blue or green.

An AND-type operation will only keep previously-displayed trajectories that are selected in all of the operation's color groups. Thus, the "Blue, Green, AND" operation would this time only keep displayed the trajectories that are selected in blue and green.

Of importance is the fact that operations and selections only affect currently displayed trajectories. This adds another level of flexibility to \revivd's filtering system, as consecutive filtering operations will be able to complement each other.

\begin{figure*}[ht!]
    \centering
    \includegraphics[width=18cm]{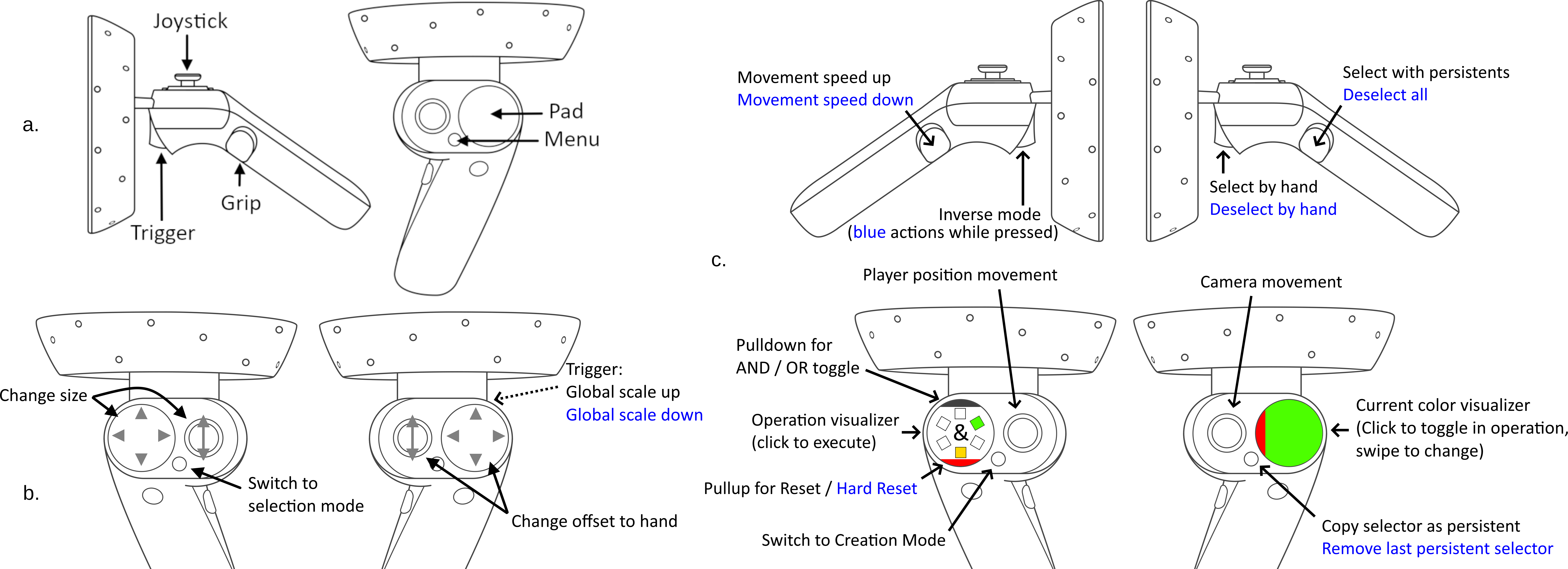}
    \caption{Description of the input of the controllers. Left: (a) Input buttons names, (b) Input mapping of "creation" mode. The controls for "creation" mode slightly differ for different selector shapes. Right (c): Input mapping of "selection" mode.}
    \label{fig:input_mapping}
\end{figure*}

\section{Controller Mapping and User Experience}

Apart from animation controls, all previously introduced features of \revivd\ are accessible directly on the controllers. We did not implement any large menu on the screen, to let the user explore the visualization with their vision unobstructed. 

The control scheme in \revivd\ was designed to give intuitive access to the previously described features, despite being restrained by the limited set of controls offered by VR controllers. Since standardized controls for VR devices are not yet available, this section is largely specific to the Samsung HMD Odyssey\footnote{\url{https://www.samsung.com/us/computing/hmd/windows-mixed-reality/}} controllers we used during the development of the project, however, most if not all of these controls could be adapted to different devices. The nomenclature of the different inputs and all of \revivd 's controls are respectively illustrated in \autoref{fig:input_mapping} (a), (b) and (c).

\subsection{Control Modes}

In order to limit the complexity of the controls, two separate control layouts are accessible and toggled at the press of a button: selection mode and creation mode. 
In selection mode, the user can select trajectories using the selectors, switch between different selector colors, place and remove persistent selectors, and create and execute filtering operations.
Creation mode is exclusive to selector modification, allowing the user to position the current selector relative to their hand and to change its dimensions; configurable parameters for each type of selector are summarized in \autoref{tab:prim_params}.

To broaden the range of possible inputs, an "inverse" trigger is used: while the left controller's trigger is pressed, some controls are modified (usually to have the opposite of their normal effect).

\subsection{Operation Controls}
Displayed on the left controller's pad are six initially blank squares, representing the six colors groups previously mentioned (\autoref{fig:input_mapping}, c). Those squares encircle a symbol which shows the operation's type: either "\&" (AND-type) or "$||$" (OR-type).

By clicking the right controller's pad, which displays the current selector's color (\autoref{fig:input_mapping}, c), this color's presence in the operation will be toggled. When a color is to be used in the operation, its corresponding square will be filled. This way, the left pad provides a complete overview of what the operation will do. A pull-down menu on the left controller's pad can be clicked to toggle the operation's type. All changes to the operation are also signaled through vibrations in the left controller, such that an experienced user will not have to look at it to know the operation's state.

Finally, the created operation can be applied to the visualization by pressing the left controller's pad (\autoref{fig:input_mapping}, c). Resetting the effects of operations requires pulling up from the bottom part of the left pad to bring up the "RESET" button, then pressing it. Finally, while the "Inverse" trigger is pressed, the left pad changes to display the word "Invert"; pressing the left pad at this moment will execute the "Invert" selection, toggling the visibility of all the paths in the dataset. This special operation is not influenced by the color groups or type of the standard operation.

\section{Implementation Notes and Optimizations}

\revivd\ is implemented in C\# using Unity\footnote{\url{https://unity.com/}}, a popular 3D real-time rendering engine traditionally used for video game development. We used Samsung's HMD Odyssey as VR equipment, with Alienware laptops to help achieve real-time rendering. Our implementation uses Valve's SteamVR library to interface with our VR equipment. \revivd 's algorithms could however be re-implemented using other software and hardware solutions.

\revivd\ is released as an open source project with permissive license. Code and documentation, as well as pre-compiled releases for Windows, are available on GitHub \ifisanonymous~[Left blank for peer review] \else (\url{https://github.com/AmigoCap/ReViVD}) \fi to facilitate use of the toolkit or the reuse of some of its parts.

While we relied on a well established stack of software and devices, an important development effort was required to achieve solid rendering performance and fast collision detection. The following sections provide a high-level overview of those algorithms and discuss the impact they had on the design of \revivd.

\subsection{Trajectories Rendering}

A crucial constraint in exploratory data analysis is to build techniques with as little latency as possible; and in VR software, maintaining a solid frame rate is critical. Indeed, bad performance will cause poor synchronisation of the video feedback to the user's movements, quickly inducing nausea; thus, \revivd\ has to be able to render the trajectories of a dataset at least 30 times per second.

\begin{figure}[t]
    \centering
    \includegraphics[width=0.7\columnwidth]{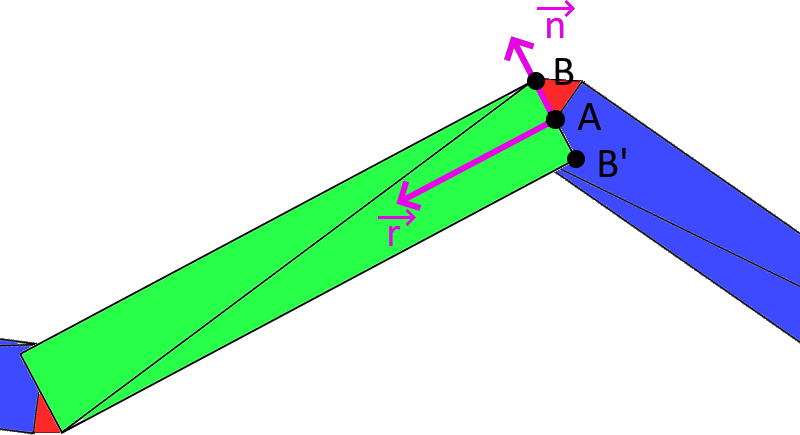}
    \caption{Structure of the procedural mesh used to render trajectories. In green and blue, ribbons; in red, junction triangles at the intersection of ribbons.}
    \label{fig:vertices}
\end{figure}

To alleviate the load on the graphics card used, we chose to render paths with a minimal amount of triangles through procedural mesh creation. Each ribbon (individual segment of a path) is made of two triangles forming a rectangle; an additional triangle is rendered between each two consecutive ribbons to soften the junction between the two rectangles. Individual meshes are procedurally created for each path by linking all of the path's ribbons. This vertex configuration is shown \autoref{fig:vertices}.

Compared to an approach where actual 3D cylinders are rendered, this method produces arguably cleaner results (the ribbons are not projections of approximated cylinders, but a fully precise 2D line) and uses fewer triangles and memory. However, it requires the ribbons to be rotated towards the camera at all time, potentially lowering the frame rate. This was done through the use of a custom HLSL (High Level Shading Language) shader. 

The shader uses the GPU to efficiently move the apparent position of the vertices of the mesh before rendering. In our case, to achieve the desired effect of a rotating ribbon facing the camera at all times, for each ribbon's end, three vertices are sent to the shader with the same position (point A on fig. \ref{fig:vertices}). The shader then ``separates" two of the three vertices along axis $\overrightarrow{n}$ to reach points B and B', which are effectively used in the rendering to give a thickness to the ribbon. Axis $\overrightarrow{n}$ itself is obtained as the normalized cross product of the ribbon's axis $\overrightarrow{r}$ and the vector from the camera to point A.

This system allows us, on  our computers equipped with Nvidia's 1080Ti (Max-Q design), to sustain an acceptable frame rate up to around 20 million on-screen atoms (constitutive points of the trajectories). The 3D rendering method was not implemented for frame rate comparison, as those results were satisfying for our purposes, while allowing drastic memory saving benefits. Indeed, compared to using hexagonal-base cylinder approximations for ribbons, this method uses 60\% less vertices (5 per junction instead of 13 per junction) and at least 75\% less triangles (3 per ribbon instead of a minimum of 13 with a single junction triangle).

\subsection{Collision Detection}

Whenever the user presses the selection trigger, the selector in their right hand becomes active, selecting and painting all the ribbons it touches in its color. Unity provides efficient methods for checking if two individual 3D shapes collide. However, a naive implementation would need to apply this check to all the ribbons in the scene and repeat this intensive operation each frame that the selector is active, making for unacceptable performance.

To alleviate this problem, space in \revivd\ is divided in small cuboid sections we call \emph{districts}. Each district keeps track of all the ribbons inside it. This partitioning allows the program to quickly get a gross approximation of which ribbons are in a specific spatial zone and to restrain the computationally expensive accurate collision detection algorithm to only those ribbons.




In order to know which districts to check for collisions, we implemented a shell-filling algorithm, particularly efficient for selections with large selectors.
\begin{enumerate}
    \item First, the center of the selector is used to determine a district necessarily touched by the selector, which we call the "center district".
    \item A line of districts stemming from the center district in an arbitrary direction is progressively checked until a district at the edge of the selector is found (which we call a "border district").
    \item A flooding algorithm starting from the found border district finds all the border districts around the selector, creating an hermetic "shell" of border districts.
    \item Starting again from the center district, a flooding algorithm finds all the districts enclosed by the shell; those districts, fully comprised within the selector, are called "inside districts".
\end{enumerate}

\begin{figure}[t]
    \centering
    \includegraphics[width=\columnwidth]{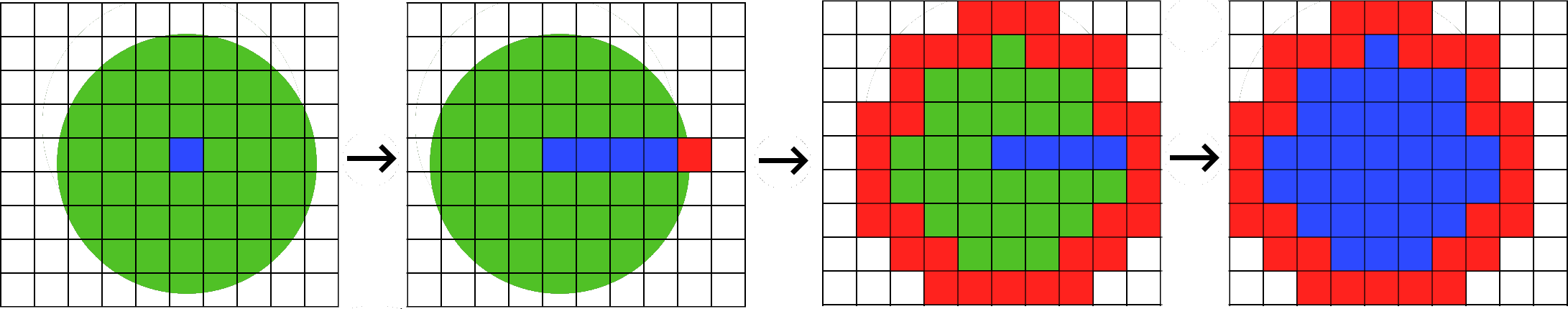}
    \caption{Four phases of the collision detection algorithm. In green, the selector; in blue, "inside"-type districts; in red, "border"-type districts.}

    \label{fig:collision}
\end{figure}

Once this has been done, the algorithm can safely assume that all ribbons pertaining to "inside" districts are touched by the selector without need for further calculations. Afterwards, only ribbons exclusive to "border" districts will be fully tested to determine if they touch the selector or not.

This algorithm, summarized \autoref{fig:collision}, essentially reduced the complexity of collision detection from $O(N^3)$ to $O(N^2)$ on very large selectors, as only the surface of the selector gets "deeply" checked.

 
\section{User Evaluation}

During the design of \revivd, we followed the Munzner~\cite{munzner_nested_2009} nested model design methodology by constantly engaging with many potential end-users from various domains, involving both human behavior trajectory recordings as well as simulations. This phase spanned over 9 months in total and helped us to validate the \emph{problem}, \emph{data abstraction} and \emph{encoding} steps of the model. This process remained highly informal, with multiple exchanges of sample datasets, questions, and screenshots of the prototype. We validated the \emph{algorithm part} of the nested model by achieving real-time frame rate for the datasets we used in our experiments.

We conducted a user evaluation for the \emph{interactions encoding} part of the nested model during formal sessions with domain-expert participants who were not involved in the design process. We sought for experts who needed to visualize large amounts of data, but with a lack of expertise in building interactive systems. In total we contacted \nbExperts experts, either from our local university or from professional networks who submitted a dataset to load into \revivd\ to us (as a technical procedure was needed to clean, format and scale the dataset). We also added static 3D models and background elements useful for context (\eg rugby pitch, map of a city/country) (see details \autoref{tab:case_properties}). We invited them to one or two evaluation sessions (depending on their availability or for follow-ups related to adding more or different datasets).

Each session started with a training procedure. After briefly presenting the tool, the HMD, and the controllers, the participants used the HMD for 30 minutes in a guided tutorial visualization of aircraft trajectories over \ifisanonymous{a country }\else{France }\fi over a day. This dataset allowed participants to use \revivd\ with a realistic exploration task in a reasonably-sized dataset with low trajectory density. The training ended with an evaluation process requiring the use of selectors and operations to solve a specific task with this dataset: selecting trajectories between two given airports. Following the tutorial, participants were able to explore their visualization and use all the capabilities of \revivd\ to filter their datasets. Each session lasted on average 60 minutes; the participants' behavior was recorded through a webcam, a screen-cast from the headset, and an automated log of their positions and interactions within the 3D environment.
Such logs enable to grasp the user activity over space and time. It also enabled us to efficiently browse video recordings to sequences of interest, \eg when multiple persistent selectors have been created.

\begin{table*}[!ht]
\centering
\small
\begin{tabular}{c|c|c|c|c|c|c|c|c|c|}
\cline{2-10}
& \multicolumn{3}{c|}{\textbf{Number of}}& \multicolumn{4}{c|}{\textbf{Attributes used for}} & \multicolumn{2}{c|}{\textbf{Context elements}} \\ \hline

\multicolumn{1}{|c|}{\textbf{Case study}} & \textbf{Trajectories} & \textbf{Atoms} & \textbf{Instants} & \textbf{X} & \textbf{Y} & \textbf{Z} & \textbf{Color}& \textbf{3D models}& \textbf{2D background} \\ \hline

\multicolumn{1}{|c|}{\gps}& $21$ & $1,118,992$ & $72,201$& x& y& \begin{tabular}[c]{@{}c@{}}Speed, Acceleration, \\\textbf{Player load},\\ Odometer, Time\end{tabular} &
\begin{tabular}[c]{@{}c@{}}\textbf{Speed}, Acceleration, \\Player load,\\ Odometer, Random\end{tabular} & Goalposts& Rugby pitch\\ \hline

\multicolumn{1}{|c|}{\propeller} & $9,440$& $738,096$ & $150$ & x& y& z & \begin{tabular}[c]{@{}c@{}}\textbf{Mach number},\\ Random\end{tabular} & \begin{tabular}[c]{@{}c@{}}Propeller,\\ Supersonic layers\end{tabular} & - \\ \hline

\multicolumn{1}{|c|}{\particles} & $10^6$ & $2.5\cdot 10^9$& $2,500$ & x& y& z & \begin{tabular}[c]{@{}c@{}}Random, Speed,\\ Acceleration, \textbf{Power}\end{tabular} & - & - \\ \hline

\multicolumn{1}{|c|}{\traffic} & $6,121$ & $4,641,944$ & $3,600$ & x& y& \begin{tabular}[c]{@{}c@{}}\textbf{Speed}, Distance, Time\end{tabular}& \begin{tabular}[c]{@{}c@{}}Speed, \textbf{Random}\end{tabular} & -& City map\\ \hline
\end{tabular}
 \vskip .5cm
\caption{Summary of the case studies by datasets properties, by count of unique trajectories, atoms (records) and instants (temporal time points) used for playback. Attributes are the ones available in the dataset for spatial mapping, for which dataset either already have a 3D structure (\propeller, \particles) or mapping of a non-spatial attribute was used (\gps, \traffic). We highlighted in \textbf{bold} the most interesting attributes during the case studies. Finally, some case studies required to add a 2D background (often a map) and a 3D model for context.}
\label{tab:case_properties}

\end{table*}


\begin{figure}[t]
    \centering
    \includegraphics[width=\columnwidth]{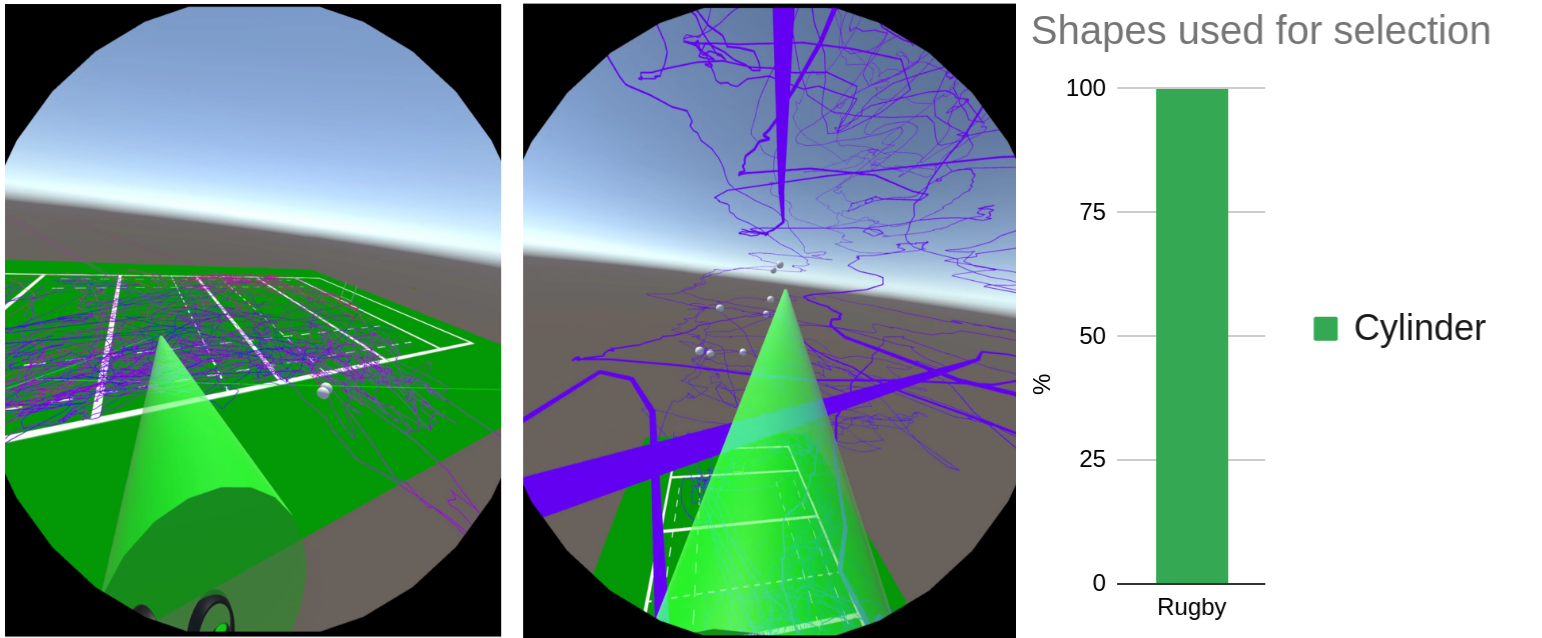}
    \caption{Case study \gps: the expert primarily used \revivd\ to visualize data from a familiar point of view and used queries to single out interesting players. He also used the animation feature to playback players position over space (x, y), and their effort over time (z).}
    \label{fig:case-gps}
\end{figure}

\subsection{Case Study \gps: Players Performance Analysis}

Sports tracking data is becoming increasingly available due to advances in wearable GPS sensor technologies. Such data provides fine-grained position of players on a pitch. We collaborated with a Rugby Team (consistently ranked in the top 5 teams of one of the major European Leagues) which constantly collects GPS data of its players and substitutes during training and games (warm-ups included). We contacted a researcher expert in rugby, with experiences as a player and trainer . He was also familiar with the tactics of the team we collaborated with and mainly used dashboards, datasheets, and video streaming to identify weaknesses in the team, which zone of the pitch was occupied the most, as well as the team behavior during an action\cite{argusa_preparation_2014}. Individual behaviors are also important to identify exhausted players needing to be substituted. We loaded the GPS dataset of a single game from the team in \revivd. As the dataset was very accurate, we only loaded 1/10 of the dataset with uniform spatial sampling. We added 3D models of goalposts and a rugby pitch map to give spatial context and references to the expert.

The first step of our domain expert was to grasp the distribution of the dataset as a standard 2D map (\autoref{fig:case-gps}, left). While this map generates an important visual clutter (due to over-plotting), the expert positioned himself in a tilted position similar to one of a spectator or a TV camera. He animated players using the playback feature and configured the paths color to reflect the players' speed. The cylinder interaction was exclusively used (\autoref{fig:case-gps}) to explore the dataset and pick a particular player (\# 14). This player's behavior was isolated, though with no particular findings. After resetting the selection, he inspected the team's collective behavior and was able to infer the positions of the forward players by looking at group placements during waves of attack or defence. He constantly changed his position around the game to adjust his point of view. For this scenario, he did not find useful to select a spatial region in particular.

The second step was to use the player load (a computed attribute representing fatigue, provided by the dataset) as the vertical coordinate of the atoms in the dataset. The results were then animated to better visualize player movement (\autoref{fig:case-gps}, middle). The resulting vertical visualization allowed the expert to quickly identify the top 3 most active players on the field. Here, cylindrical selectors were often used as pointing rods to explain parts of the game or point to particular players without using their selecting ability.


\begin{figure}[h]
    \centering
    \includegraphics[width=\columnwidth]{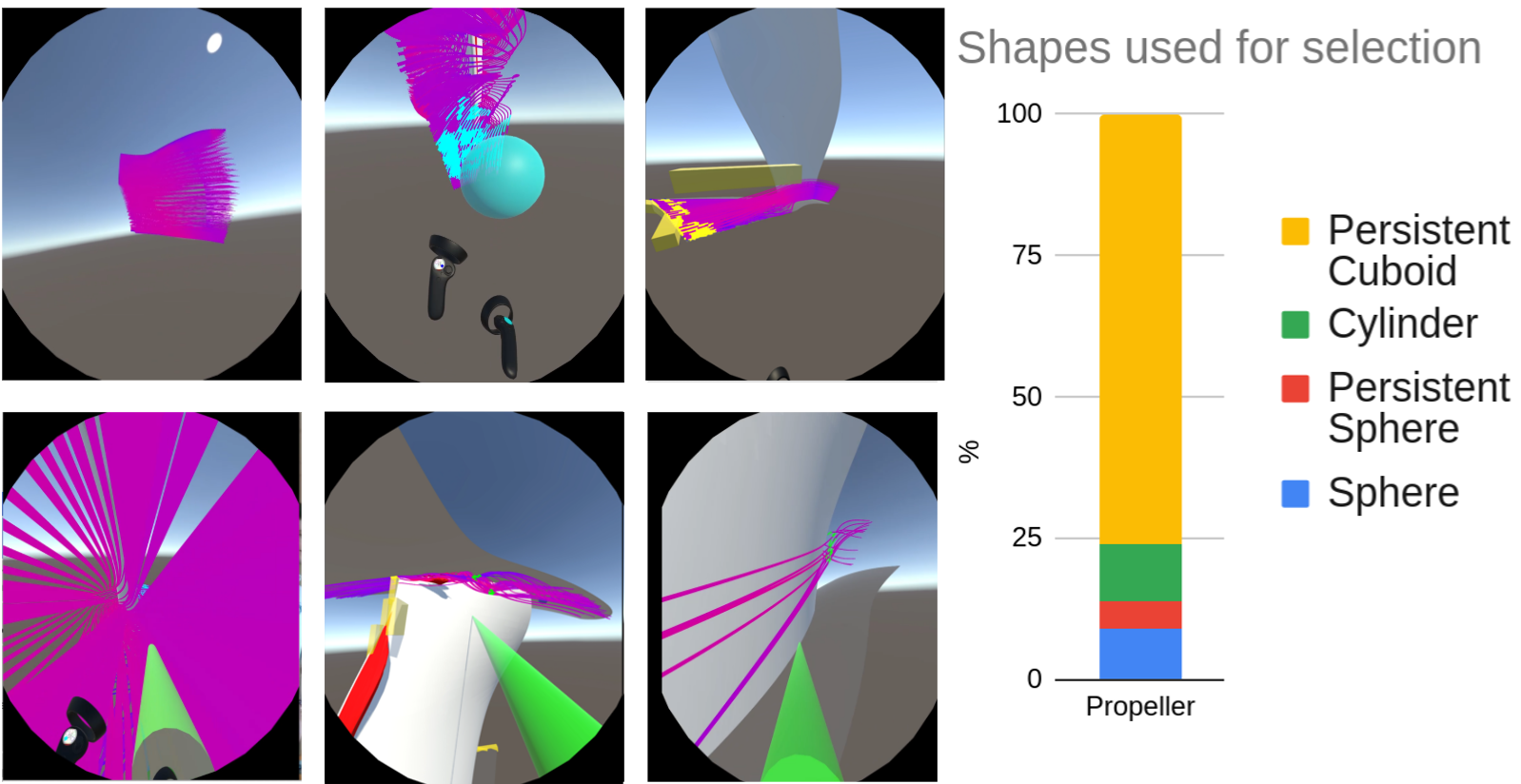}
    \caption{Case study \propeller: the experts managed to quickly select streamlines of interest around a propeller. Selections were refined to single out streamlines involved in the creation of the vortex and separated flow areas.}
    \label{fig:case-turbulence}
\end{figure}

\subsection{Case Study \propeller: Flow Simulation Exploration}

Our colleagues from the Engineering department of our local university provided us with a numerical simulation of turbulence phenomena around a propeller over time {\ifisanonymous \else \cite{brandstetter_project_2019}\fi}. After discussions, we selected a data subset from a numerical simulation of $9,440$ streamlines over a maximum of $150$ instants, resulting in a total of $738,096$ atoms. Those streamlines were localized around a propeller currently studied by these experts. Such a structure generates regions of interest such as a vortex and separated flow regions. They typically wanted to know if the streamlines involved in the vortex were from the same departure areas or not, and to study their time evolution. They also sought to identify supersonic regions in the flow. Experts knew interesting areas in advance and were seeking for a quick way to explore and visualize their dataset, as well as the possibility of communicating their research to either students or the general public in intuitive video format.
They currently use Mathwork's Matlab\footnote{\url{https://www.mathworks.com/products/matlab.html}} and Kitware's ParaView to run their simulation and visualize the results.

As streamlines are already by nature 3D data, their representation is straightforward. We tested two color strategies: coloring by Mach number to highlight the supersonic area, or using random colors to study the streamlines separately. 3D models of the propeller and of the supersonic layers completed the visualization.

One of the experts started his exploration with an overview of the trajectories and then selected streamlines from departure locations he knew would hold streamlines involved in vortex. Then, he changed his point of view to refine the selection using the cylinder shape (due also to a preference for what he called "a lightsaber”). By using another color, he could see the newly selected trajectories as a subset of previous selections. \autoref{fig:case-turbulence} illustrates this step-by-step process up to the final selection. The expert then spent some time explaining the physics foundations behind this phenomena using the cylinder to point to and describe the behavior of the selected streamlines.

Another expert focused on mostly the same phenomena, but approached the selection process differently, using many wall-shaped persistent cuboid selectors to filter streams by their spatial origin.

\subsection{Case study \particles: 3D Particles in a Turbulent Flow}

Other colleagues from the Engineering department provided us with a simulation of 3D particles in an isotropic and turbulent flow \cite{naso_multiscale_2019}. Due to the large number of particles, they face visualization and exploration problems with this dataset. Surprisingly, they had never been able to simultaneously visualize a correct amount of particles even though it is an important aspect of their research work. Indeed, they wanted to explore their turbulent flow to develop an intuition on data and results and confirm predictions on the evolution of attributes according to the shape of the trajectory. They currently use Mathwork's Matlab to visualize these particle trajectories, but this tool only allows them to extract and visualize some parts of the dataset. We met these colleagues to discuss their dataset and visualization problems. This allowed us to identify an interesting subset of data that we enriched by computing the speed, acceleration, and power attributes for all the particles. We ended up with a dataset describing the evolution of a million particles in a turbulent flow over $2,500$ instants, which means $2,500,000,000$ records in total.

\begin{figure}[t]
    \centering
    \includegraphics[width=\columnwidth]{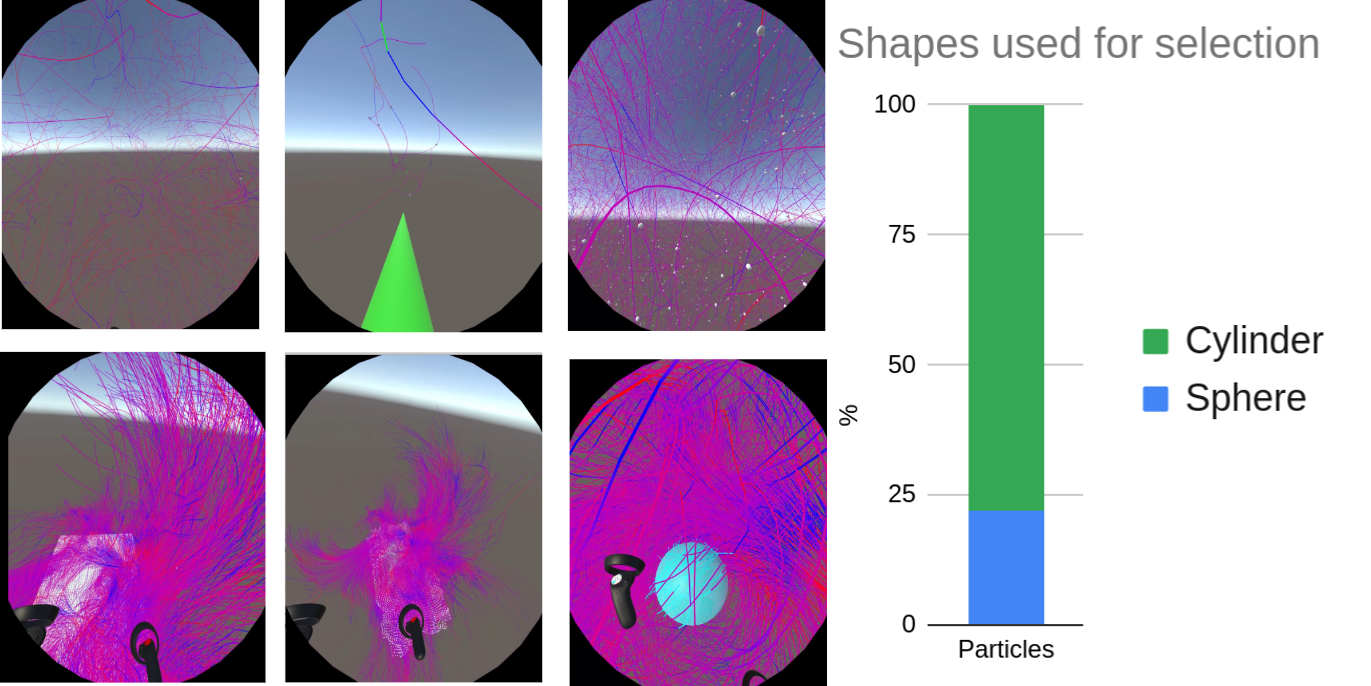}
    \caption{Case study \particles: the experts studied the changes in power of the particles relative to the shape of the trajectories.}
    \label{fig:case-particle}
\end{figure}

The experts started exploring from within the particles' trajectories (\autoref{fig:case-particle}). The speed, acceleration, and power attributes were used to color the trajectories along with random colors for each particle depending on the focus of the expert. As \revivd\ could not display all the points available in the dataset, we had to sample the dataset. A standard sampling example, used several times, was a visualization of $20,000$ particles chosen randomly in the dataset between the instants $500$ and $1,500$ amounting to $20,000,000$ atoms. No context elements such as 3D models or 2D background were added to this visualization. The experts were primarily interested in identifying changes in the power received by the particle from the surrounding flow.

\subsection{Case Study \traffic: Traffic Flow Simulation}

Understanding flows and movements of vehicles in large cities is important to apprehend the mobility of the residents. This data can lead to reductions in contamination and improvements to the traffic and living conditions. However, such data is usually difficult to collect as not all cars share their coordinates in real time. We collaborated with a Research laboratory specialized in simulation traffic data, in particular with one of their latest dataset over \city \cite{paipuri_validation_2019}. It contained $6,121$ trajectories of vehicles simulated over $3,600$ instants (1 hour of traffic simulation), making a total of $6,641,944$ atoms (or records). Such a dataset enables following individual cars with a very high (simulated) spatial resolution. The experts were interested in identifying congestion areas and sources of traffic that can create it. They also needed a visualization tool for their results to quickly validate the parameters of the simulation. They currently use Ifsttar's Symuvia\footnote{\url{https://www.licit.ifsttar.fr/linstitut/cosys/laboratoires/licit-ifsttar/plateformes/symuvia/}}, a specialized simulation tool for traffic with 2D visualizations of the results to generate this dataset, as well as Mathworks' Matlab for all the post-processing and analysis of data.

\begin{figure}[t]
    \centering
    \includegraphics[width=.8\columnwidth]{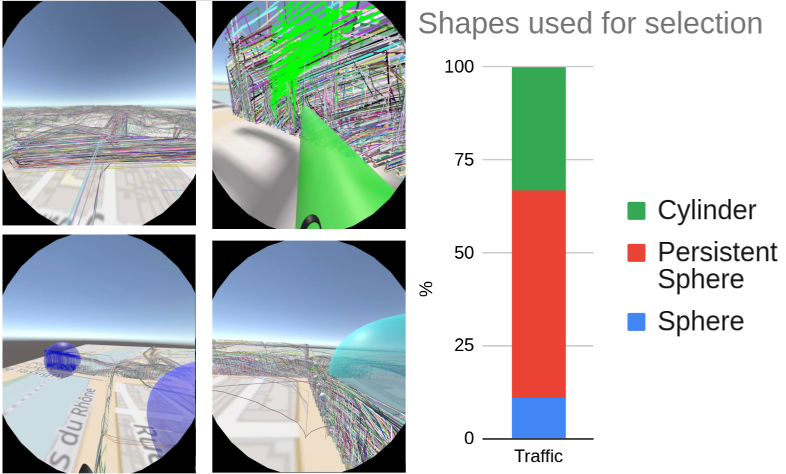}
    \caption{Case study \traffic: the expert studied the vehicle trajectories likely to create congestion through the main critical junctions of the city.}
    \label{fig:case-traffic}
\end{figure}

The expert started exploring the dataset and first used a cylindrical, saber-like selector to extract the vehicles passing through a critical junction of the city and study their global trajectory (\autoref{fig:case-traffic}). The second step was to reset the selected trajectories and use persistent spheres, strategically placed on main junctions of a district of the city, to extract only these trajectories. He then used the sphere-drop functionality to only animate the vehicles on this selection and study the congested junctions. This simulation traffic dataset was intrinsically made of 2D trajectories on a plane, so we chose the speed of the vehicles as a third dimension. During the evaluation session, the coloration of the trajectories was either unique for each trajectory or representative of the speed of the vehicle. Despite the density of points, \revivd\ was able to handle the full visualization and all the vehicle trajectories were displayed. We added the map of {\ifisanonymous the city \else Lyon \fi} as a background to give spatial references to the expert.

\section{Discussion}

\begin{figure}[h]
    \centering
    \includegraphics[width=\columnwidth]{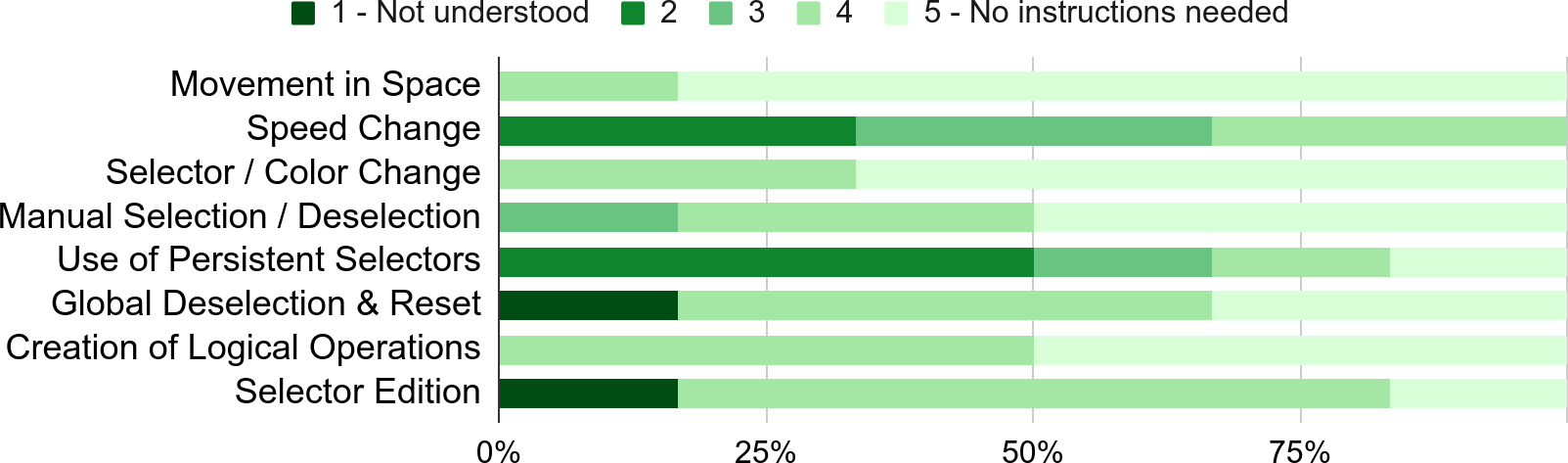}
    \caption{Evaluation of the intuitiveness of the command mapping by the \nbExperts participants on a Likert scale ranging from 1 to 5 (with 5 being the most intuitive).}
    \label{fig:likert}
\end{figure}

We collected post-study feedback from users regarding the intuitiveness of the tool. Participants filled a questionnaire with a Likert scale ranging from 1 to 5 (with 1 corresponding to the less intuitive command mappings) (\autoref{fig:likert}). The most intuitive commands were the movement in space using the Joystick, change of selector / color, and creation of Boolean operands. Change of speed, use of persistent selectors, and manual selection/deselection of queries were found quite intuitive. The reset command and creation mode were also found quite useful by participants, but one was not able either to use nor remember those commands. Performance was rated very good, but it could be improved to load larger datasets, mainly by refactoring the code to use a Data-Oriented Programming (DOP) approach to facilitate multi-threading, as well as moving collision detection computation to the graphics card.

The expert of the case study \gps\ raised some concerns about the additional value of VR in the exploration of the dataset. Indeed, most of the operations that he had done, could have been done in a desktop environment. Playback of the players could be achieved through video streaming of the game, and the identification of the most exhausted players through a dashboard or the use of datasheets. Nevertheless, the expert told us that \revivd\ and VR might bring something new to the field if other attributes (\eg heart rate) or automatic recognition of similar trajectory patterns were available. This would for instance allow for comparison of data between similar players, as well as a visualization of the players' effort during specific actions (\eg rucks). 

Further exchange with the experts allowed us to find two other flaws of \revivd\ in its current state. First, even though \revivd\ supports very early exploratory analysis, preliminary work is still needed to prepare the datasets. During our user evaluation, we pre-processed some datasets to either filter or extract simulation samples as the entire dataset was too accurate in terms of number of records to be fully displayed at high frame rate. We also added domain-knowledge (\eg 3D models, 2D background) to contextualize the exploration process. Those steps were purely technical and were achieved after multiple exchanges with domain experts. This was, however, partially addressed in later development stages, as the latest version of \revivd\ facilitates this process by letting experts load their own trajectory datasets through a \texttt{.json} configuration file.

Secondly, exploration of 3D structures requires easy access to hidden, additional attributes, which \revivd\ can for now only show after a reload of the entire visualization. During the case studies, an operator assisted the user to switch z-axis mapping and get details on the datasets information (\eg names of moving entities) as those features are inaccessible in real-time in \revivd. Adding those features has a design implication that requires further research. Such features are needed to fully support analytical processes~\cite{chandler_immersive_2015}, along with additional standard interactions such as sorting, deriving, or aggregating values.

One of our main research perspectives is to investigate more diverse shapes for selectors, beyond the three basic shapes we implemented. According to~\cite{wills_selection:_1996}, two directions could be explored: \emph{custom} shapes and \emph{data-dependent} shapes. Custom shapes operate on the geometry of the shape, independently from the data. CAD (Computer Aided Design) tools (\eg Autocad\footnote{\url{https://www.autodesk.com/products/autocad/overview}}) already support building such shapes, along with operations such as the subtraction of a shape by another. Extrusion could be a promising operation to build custom shapes using the 3D models of the scenes, as we have seen those greatly influence selections. For instance, experts of the \propeller\ case study could create a contour of the propeller, by subtracting a cube from the propeller shape instead of adding small spheres together to re-construct this contour. Gestures could be used to create such shapes, and an extensive body of work is already investigated this area (\eg~\cite{vinayak_shape-it-up:_2013}).

Data-dependent shapes are driven by the data distribution, so that the shape selects groups of data points optimally using the density of the surrounding region~\cite{yu_efficient_2012} or non-spatial attributes~\cite{jackson_force_2012}. Combining shapes with data beyond selections has a very high potential. For example, remote selection of trajectories could be facilitated by encoding the density as a visual property of the shape. Similarly, density can be encoded in haptic feedback~\cite{prouzeau_scaptics_2019}) to access cluttered trajectories.



    



\section{Conclusion}

We presented \revivd, a tool for exploring and filtering large trajectory-based datasets using virtual reality. It extends the design space of volumetric probes by introducing a set of configurable 3D shapes and logical operations for use in intuitive selection. We evaluated its use case with domain-experts who sought to quickly grasp an overview of their dataset, as well as to extract particular regions or trajectories of interest. Our results show the ease of use and expressiveness of the 3D geometric shapes implemented in \revivd\ as selectors. \revivd\ was found to be particularly useful for progressively refining selections from extensive trajectory datasets to small subsets of outlying behavior. It also demonstrated its use as a powerful tool for conveying the structure of normally abstract datasets for communication purposes.

\acknowledgments{
\ifisanonymous
[Left blank for peer review]
\else
We thank all the experts who shared their data to visualize with \revivd\ and provided helpful comments and advice on our visualization tool.
This project was partially supported by the M2i on Urban Mobility funded by the French Agency for Durable Development (ADEME) and by a BQR (Bonus Qualité Recherche) project funded by the École Centrale de Lyon.
\fi
}

\bibliographystyle{abbrv-doi}

\bibliography{template}
\end{document}